\documentclass{article}
\usepackage{spconf,amsmath,graphicx}
\usepackage{amssymb,amsthm,amsmath}
\usepackage{color,multicol}

\title{Nonparametric methods for detecting change in Multitemporal SAR/PolSAR Satellite Data}
\name{Rodney Fonseca$^{\dagger}$ \qquad Guilherme Ludwig$^{\dagger}$ \qquad Michel Montoril$^{\star}$ \qquad Alu\'{i}sio Pinheiro$^{\dagger}$
\thanks{The authors acknowledge support from S\~{a}o Paulo Research Foundation (FAPESP), grant 2018/04654-9. R. Fonseca acknowledges grant 2016/24469-6 (FAPESP). G. Ludwig acknowledges grant 2019/03517-0 (FAPESP). A. Pinheiro acknowledges grant 309230/2017-9 (CNPq). The Forest data set was kindly provided by Abdourrahmane Atto.}
}
\address{
\hspace{2em}$^{\dagger}$ University of Campinas \qquad
\hspace{4em}$^{\star}$ Federal University of S\~{a}o Carlos\\
\hspace{0em} Statistics Department, Campinas, Brazil \qquad \hspace{0em}Statistics Department, S\~{a}o Carlos, Brazil\\
}
\begin{document}
%
\maketitle
\begin{abstract}
We employ nonparametric statistical procedures to analyse multitemporal SAR/PolSAR satellite images. The aim is two-fold. We seek parsimony in data representation as well as efficient change detection. For these, wavelets and geostatistical analyses are applied to the images (Morettin et al., 2017; Krainski et al., 2018). Following this representation, the dimension of the underlying generating process is estimated (Fonseca and Pinheiro, 2019), and a set of multivariate characteristics is extracted. Change-points are then detected via wavelets (Montoril et al., 2019).
\end{abstract}
\section{Introduction}
\label{sec:intro}

In order to analyze time series composed of satellite images, \cite{atto2012multidate} propose	 a method to compute a divergence measure using features of images along time. It consists in decomposing the images on wavelet bases, which provides subband coefficients for each time point. Later, parametric models are fitted for the coefficient's distribution, and they are used to compute a Kullback-Leibler (KL) divergence measure for images on distinct time points. These measurements are used to form a matrix called multi-date divergence matrix (MDDM) that is used to analyze how the image varies through time.

In this paper we propose novel ways of detecting change points in satellite image time series taking as reference the idea of \cite{atto2012multidate}, but employing nonparametric estimates of the density functions. Applying wavelets as well as nonparametric methods even further in the analysis, we estimate the square root of the density function of coefficients from the image decomposition, and refine these estimates using a functional approach that accounts for the dimension of the subspace generating the time series of functions. This method provides tools for two types of change point analysis: through wavelet coefficients, that can be used directly to compute the Hellinger distance as divergence measure in order to form a MDDM as the original idea of \cite{atto2012multidate}; or through time series of loadings that can be used in a mixture problem where one of the classes is taken as representing changed time points. These methods are illustrated with an application to a time series of forest images in a region between Brazil and the French Guiana.

\section{Multi-date divergence matrix}

In this section we describe how the method proposed by \cite{atto2012multidate} may be used to detect time points in which changes occur in temporal images. Employing the same notation as the authors, let us denote a time series of $M$ images as $\pmb{\mathcal{I}}=\{\mathcal{I}_m;m=1,\ldots,M\}$. The first step in the analysis is to decompose the images on wavelet basis, which for stationary wavelet transform (SWT) can be represented as
\begin{align}
\mathcal{I}_m = &\sum_{k_1,k_2}\langle\mathcal{I}_m,\Phi_{J,[k_1,k_2]}\rangle\Phi_{J,[k_1,k_2]} +\nonumber\\
&\sum_{j=1}^{J}\sum_{k_1,k_2,n_1,n_2}\langle\mathcal{I}_m,\Psi_{j,[n_1,n_2],[k_1,k_2]}\rangle\Psi_{j,[n_1,n_2],[k_1,k_2]},
\label{E:swt_image}
\end{align}
where $n_1,n_2\in\{0,1\}\setminus\{0,0\}$, $\Phi_{J,[k_1,k_2]}$ and $\Psi_{j,[n_1,n_2],[k_1,k_2]}$ represent translated and scaled versions of a scaling function $\Phi$ and detail function $\Psi$. The indexes $n_1$ and $n_2$ indicate the four different subbands obtained, usually denoted by their corresponding region on $\mathcal{I}_m$: approximation, horizontal, vertical and diagonal subbands. The wavelet coefficients shall be denoted as $c_{J,[0,0],[k_1,k_2]}[\mathcal{I}_m]=\langle\mathcal{I}_m,\Phi_{J,[k_1,k_2]}\rangle$ for approximation coefficients and $c_{j,[n_1,n_2],[k_1,k_2]}[\mathcal{I}_m]=\langle\mathcal{I}_m,\Psi_{j,[n_1,n_2],[k_1,k_2]}\rangle$ for detail coefficients.

In the second part of the analysis, the coefficients on each subband are vectorized, say $\mathbf{c}_{[0,0]}[m] = \left\{c_{J,[0,0],[k_1,k_2]}[\mathcal{I}_m]\right.$; $\left. k_1,k_2\right\}$ and $\mathbf{c}_{[n_1,n_2]}[m] = \left\{c_{j,[n_1,n_2],[k_1,k_2]}[\mathcal{I}_m];j,k_1,k_2\right\}$, and a density function is estimated for each of these vectors, which \cite{atto2012multidate} proposed to be done with parametric models. Let us denote the associated estimated models as $f_{[n_1,n_2]}[m]$, $n_1,n_2\in\{0,1\}$.

The process above is applied on all images $\mathcal{I}_m$, and then a divergence measure between each pair of images is computed using the Kullback-Leibler distance for the models in each subband:
\begin{align*}
\mathcal{K}(m,l) = \sum_{n_1,n_2}KL(f_{[n_1,n_2]}[m],f_{[n_1,n_2]}[l]),
\end{align*}
where $m,l\in\{1,\ldots,M\}, m\neq l,$ and $KL$ denotes the Kullback-Leibler distance. Then, a symmetric matrix is built with upper triangular part given by $[\mathcal{K}(m,l)]_{1\leq m < l \leq M}$. This matrix is used to evaluate how the image changed throughout the $M$ time points and helping identify when the most expressive changes occurred. This matrix is called multi-date divergence matrix. For instance, the first row and first column of the MDDM compares the first image $\mathcal{I}_1$ with all the other images, and should display an increasing pattern when there are cumulative modifications in the investigated region.

\section{Nonparametric density estimation for the MDDM method}
\label{sec:nonparametric_MDDM}

We propose a nonparametric approach for estimating the density function of the subband coefficients' distributions when computing a MDDM matrix. We follow the idea of \cite{pinheiro1997estimating} and compute the wavelet representation of the square root of the density function $\sqrt{f}$ instead of $f$ itself. Among the advantages of doing so, we can highlight the fact that these estimates of $f$ are always non-negative, since $\hat{f} = (\widehat{f^{1/2}})^2$, and the integral of an estimated $f$ can be easily set to be one, by making the wavelet coefficients corresponding to $\widehat{f^{1/2}}$ have norm one. Additionally, using the Hellinger distance as divergence measure, we can use the wavelet coefficients of $\widehat{f^{1/2}}$ directly to compute it:
\begin{align*}
He(f,g) &= \left(\frac{1}{2}\int\left\{\sqrt{f(x)}-\sqrt{g(x)}\right\}^2dx\right)^{1/2}\\
&= ||\pmb{\alpha}_{f^{1/2}} - \pmb{\alpha}_{g^{1/2}}||_{\ell_2},
\end{align*}
where $\pmb{\alpha}_{f^{1/2}}$ and $\pmb{\alpha}_{g^{1/2}}$ denote the wavelet coefficients of the density functions $f$ and $g$ respectively. Moreover, fast computation and sparse representations are presented.

\subsection{Functional dimension estimation}

After the stage of density function estimation, we have a time series of square root of density functions corresponding to each level of detail/approximation of the wavelet transform. Assuming that these time series form a stationary process, since the square root of density function is square integrable, we can apply the method discussed by \cite{fonseca2020wavelet} to estimate the dimension of the subspace generating each curve time series. We obtain an improved estimate of the curves since unexplained temporal structures are removed in the functional representation, which is then described by a finite set of fixed eigenfunctions and a vector time series of loadings. Besides, it allows us to make predictions of future density functions, using the loadings time series.

Let us denote an observed curve time series by $\{f_m;t=1,\ldots,M\}$, $\bar{f}=M^{-1}\sum_{m=1}^M f_m$, and assume they can be represented in wavelet bases:
\begin{align*}
f_m(x) - \bar{f}(x) = \sum_{j=1}^{J}c_j^t\phi_j(x),
\end{align*}
where $\phi$ denote wavelet functions of some basis and $J$ is the number of coefficients employed in this representation, which in practice is finite. The dimension of the subspace generating this time series can be estimated evaluating which eigenvalues of a matrix $\mathbf{D}$ with elements
\begin{align*}
D_{j,j'} = \frac{1}{(M-p)^2}\sum_{k=1}^{M}\sum_{r,s=1}^{M-p}\sum_{l=1}^{J}c_{j}^{r}c_{j'}^{s}c_{l}^{r+k}c_{l}^{s+k},
\end{align*}
$j,j'=1,\ldots,J$, are significantly greater than zero, where $p$ can be taken as a small positive integer. The eigenvalues are tested sequentially through bootstrap tests until the first one that is not significantly zero is found, say the $(\hat{d}+1)$th eigenvalue. In this case, we take $\hat{d}$ as the estimated dimension of the process. The first $\hat{d}$ columns of $\mathbf{D}$ have wavelet coefficients of eigenfunctions $h_1,\ldots,h_{\hat{d}}$ that can be used to represent the curve time series, whose estimate can be reconstructed by applying the inverse wavelet transformation on these coefficients to obtain
\begin{align*}
\hat{f}_m(x) = \bar{f}(x) + \sum_{k=1}^{\hat{d}}\eta_{k,m} h_k(x),
\end{align*}
where the loadings $\pmb{\eta}_m=(\eta_{1,m},\ldots,\eta_{\hat{d},m})^{\top}$ are responsible for the temporal dynamics of $\hat{f}_m$.

In summary, the new algorithm to compute a fully nonparametric MDDM for a time series of images $\pmb{\mathcal{I}}=\{\mathcal{I}_m;m=1,\ldots,M\}$ consists in performing the following steps for all pairs $(m,l)$ such that $1\leq m<l\leq M$:
\begin{enumerate}
\item Decompose the images $\mathcal{I}_m$ using a bi-dimensional discrete wavelet transform (DWT);
\item Using the vectorized wavelet coefficients of the previous step, $\mathbf{c}_{[n_1,n_2]}[m]$, $n_1,n_2\in\{0,1\}$, apply the wavelet method to estimate the square root density corresponding to each subband. It results in vectors of wavelet coefficients $\pmb{\alpha}_{[n_1,n_2]}[m]$, $m=1,\ldots,M$, corresponding to a curve time series.
\item Estimate the dimension of the subspace generating the curves corresponding to $\pmb{\alpha}_{[n_1,n_2]}[m]$, $m=1,\ldots,M$, and take the functions generated with the estimated dimension as final estimates of the square root of the density curves.
\item Using the coefficients corresponding to these curve estimates at times $m$ and $l$, say $\hat{\pmb{\alpha}}_{[n_1,n_2]}[m]$ and $\hat{\pmb{\alpha}}_{[n_1,n_2]}[l]$, compute the Hellinger divergence measure as
\begin{align*}
\mathcal{K}(m,l) &= \mathcal{K}(l,m) =\\
&\sum_{n_1,n_2}\left\Vert \frac{\hat{\pmb{\alpha}}_{[n_1,n_2]}[m]}{\Vert\hat{\pmb{\alpha}}_{[n_1,n_2]}[m]\Vert_{\ell_2}}  - \frac{\hat{\pmb{\alpha}}_{[n_1,n_2]}[l]}{\Vert\hat{\pmb{\alpha}}_{[n_1,n_2]}[l]\Vert_{\ell_2}} \right\Vert_{\ell_2},
\end{align*}
where the normalization is done to guarantee the corresponding estimates of the density functions integrate one.
\item Repeating steps (1)-(4) for all pairs $(m,l)$ such that $1\leq m<l\leq M$ gives the MDDM: $\{\mathcal{K}(i,j)\}_{1\leq i,j\leq M-1}$.
\end{enumerate}

\section{Nonparametric mixture problem}
\label{sec:mixture_method}

In this section we discuss a different approach to identify change points with wavelet techniques. We shall assume that the time series of loadings obtained in the functional decomposition come from two different populations, which are mixed according to some mixture function that varies through time. We intend to estimate such a function using the method proposed by \cite{montoril2019wavelet}. The authors propose a wavelet method to deal with this mixture problem.

In their setup, \cite{montoril2019wavelet} consider that an observation of interest might come from two random variables $U_t$ or $V_t$, but that we observe only
\begin{align*}
Y_t = Z_t U_t + (1 - Z_t)V_t, \quad t\in[0,1],
\end{align*}
where $Z_t$ is a random variable following a Bernoulli distribution with parameter $\rho(t)$, $0\leq\rho(t)\leq 1$, which is called a mixture function. Assuming that $U_t$ and $V_t$ have known means $\mu_U$ and $\mu_V$, respectively, under some other assumptions, $\rho(t)$ can be seen as the expectation of $W_t = (Y_t - \mu_V)/(\mu_U - \mu_V)$. Hence, the mixture function can be estimated as a heteroscedastic regression problem involving the observed variables $W_{t_1}, \ldots, W_{t_n}$ and the time points $t_1,\ldots,t_n$, which are analyzed through wavelet methods by \cite{montoril2019wavelet}.

Our idea is to consider that loadings $\pmb{\eta}_m$, $m=1,\ldots,n$, come from two random variables, where one of them is related to images with abrupt changes, which we could identify by analyzing a mixture function estimate. The loadings are separated in two groups, whose sample means play the roles of $\mu_U$ and $\mu_V$, and equally spaced time points are considered to apply to the mixture method. Since we expect a few time points with abrupt changes on the images, the application is expected to lead to a mixture function with bumps, a case where wavelet methods are known to perform well and can provide better descriptions about changes points in the image time series.

\section{Kriging method}
\label{sec:kriging_method}

Kriging is a geostatistics method for interpolating data values indexed spatially under the assumption that data are samples of an underlying Gaussian random field \cite{cressie1993statistics,krainski2018advanced}. Kriging can be used for smoothing and pre-processing remote sensing data under suitable conditions \cite{stein1998integrating}. Let $x_1, x_2, \ldots, x_n$ represent the set of spatial coordinates corresponding to each image pixel, and $\mathcal{I}_m = (\mathcal{I}_m(x_1), \ldots, \mathcal{I}_m(x_n))^t$ an image at time $m.$ Assume that $\mathcal{I}_m$ is an intrinsic stationary isotropic random field, such that $\mathbb{E}(\mathcal{I}_m(x_i)) = \mu_m$  for $i = 1, \ldots, n,$ and \[\mathbb{E}\left[(\mathcal{I}_m(x_i) - \mathcal{I}_m(x_j))^2\right] = 2 \tau^2 + 2 \sigma^2 \left[1 - \rho_{\boldsymbol\theta}(\|x_i - x_j\|) \right]\] for $i = 1, \ldots, n,$ and $j=1, \ldots, n;$ and $\rho_{\boldsymbol\theta}$ a positive definite function with $\rho_{\boldsymbol\theta}(0)=1$ for all parameters $\boldsymbol\theta.$ Determining the covariance model $\rho_{\boldsymbol\theta}$ involves examining the empirical variogram of the images; we refer the reader to \cite{cressie1993statistics} and \cite{stein1999interpolation} for discussion. We assume for simplicity that $\tau^2, \sigma^2, \rho_{\boldsymbol\theta}$ and $\boldsymbol\theta$ are the same across all images $m = 1, \ldots, M.$

The ordinary kriging of $\mathcal{I}_m$ at a site $x_0$ is 
\begin{equation}
\mathcal{I}_m(x_0) = \hat{\mu}_m+\hat{\mathbf{c}}^t\hat{\boldsymbol\Sigma}^{-1}(\mathcal{I}_m - \hat{\mu}_m \mathbf{1}),
\label{eq:krig}
\end{equation} where $\mathbf{1}$ is a $n \times 1$ vector of ones, $\hat{\mu}_m = \mathbf{1}^t \hat{\boldsymbol\Sigma}^{-1} \mathcal{I}_m / \mathbf{1}^t \hat{\boldsymbol\Sigma}^{-1}\mathbf{1},$ $\hat{\mathbf{c}}^t = \hat{\sigma}^2(\rho_{\hat{\boldsymbol\theta}}(\|x_0 - x_1\|), \cdots, \rho_{\hat{\boldsymbol\theta}}(\|x_0 - x_n\|))$ and that $\boldsymbol\Sigma$ is a matrix with entries $\Sigma_{ij} = \hat{\tau}^2 1_{\{i=j\}} +  \hat{\sigma}^2 \rho_{\hat{\boldsymbol\theta}}(\|x_i - x_j\|)$ and $1_{A}$ is the indicator function of the set $A.$ Here $\hat{\tau}^2, \hat{\sigma}^2$ and $\hat{\boldsymbol\theta}$ are minimum contrast estimators, which minimize a distance metric between the empirical and theoretical variogram \cite{cressie1993statistics}.

Since the expression \eqref{eq:krig} involves a matrix inversion, it can be prohibitive to compute unless $\boldsymbol\Sigma$ is sparse, or if a tapering function is used to find a sparse approximation to the kriging predictor \cite{furrer2006covariance}. For the case study, following an inspection of the empirical variograms, we choose an exponential covariance model with a Wendland covariance taper.

\section{Application}

In this section we apply the nonparametric MDDM method to analyze a time series of 87 satellite images of a Tropical Forest region in the border of Brazil and the French Guiana. The application of MDDM for such time series might be useful to verify changes happening in this area, such as floods or dry periods.

As it is common on analysis of satellite images, we shall work with the logarithm of the observed images. This way we consider the presence of multiplicative noise, like the speckle that affects SAR images. Afterwards, we shall employ a smoothing method on the images to reduce the presence of noise and use these smoothed images on the MDDM method described on section \ref{sec:nonparametric_MDDM}. The smoothing method considered is a wavelet thresholding applied to images, where it is decomposed on bi-dimensional DWT, then a soft thresholding is applied on detail coefficients and the image is recovered with an inverse DWT. We also used the kriging method to perform a pre-smoothing of the images before computing the MDDM, which demands more computational time but can offer alternative results. The following figures allow us to compare the results in the cases when kriging is applied or not.

\begin{figure}[htb]
\begin{minipage}[b]{1.0\linewidth}
  \centering
  \centerline{\includegraphics[width=8.5cm]{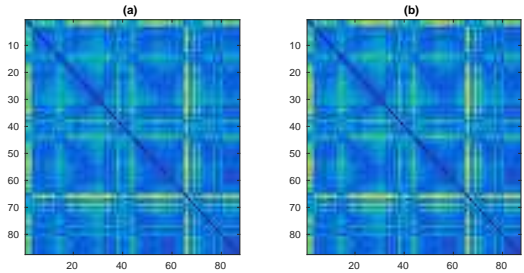}}
\end{minipage}
\caption{MDDM corresponding to the forest image time series without kriging pre-smoothing (left) and with kriging pre-smoothing (right).}
\label{fig:MDDMs}
\end{figure}

On Figure \ref{fig:MDDMs} is shown the MDDMs obtained for this time series with and without kriging. We can observe that images are very similar, displaying time points around 1 and 65 as the most different compared with all the others

\begin{figure}[htb]
\begin{minipage}[b]{1.0\linewidth}
  \centering
  \centerline{\includegraphics[width=8.0cm]{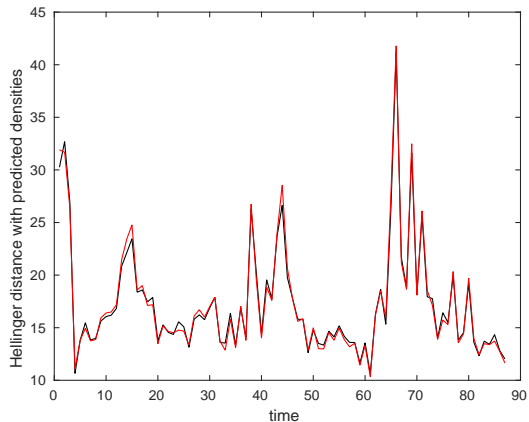}}
\end{minipage}
\caption{Hellinger distance corresponding to observed images and a predicted image, which had the wavelet coefficient’s densities predicted without kriging pre-smoothing (black) and with kriging pre-smoothing (red).}
\label{fig:predHellingerDist}
\end{figure}

Using the loadings obtained from the representation of the curve time series in finite basis, we performed a prediction of future observations of these loadings, which were used to obtain estimates of future density functions. That allowed us to estimate the Hellinger distance between a predicted image and the images already observed, as is presented in Figure \ref{fig:predHellingerDist}. The pattern in this figure is similar for both cases when kriging is applied or not, with the largest differences being observed for the time points 2, 15, 44 and 66, and having in general small differences in comparison with other time points. This might be an indicative that the future image of this time series is likely to be more similar to the mean image rather than present expressive changes compared to it.

\begin{figure}[htb]
\begin{minipage}[b]{1.0\linewidth}
  \centering
  \centerline{\includegraphics[width=8.5cm]{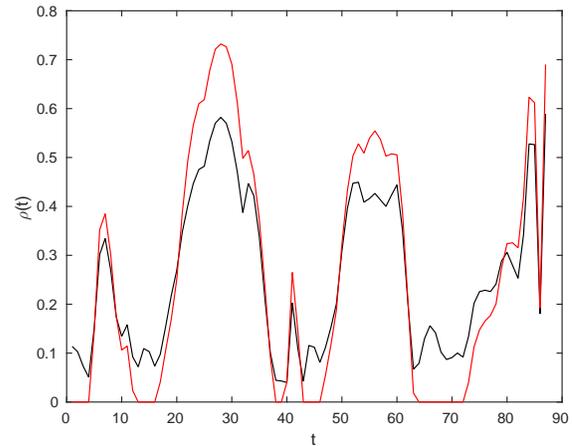}}
\end{minipage}
\caption{Mean mixture function corresponding of loadings corresponding to approximation coefficients and coarser details, without kriging pre-smoothing (black) and with kriging pre-smoothing (red).}
\label{fig:mixtureProbs}
\end{figure}

We also applied the mixture method discussed in Section \ref{sec:mixture_method} on the loadings obtained from the functional representation corresponding to approximation and coarser detail coefficients. The mean mixture functions of these loadings are presented in Figure \ref{fig:mixtureProbs}. The valleys observed in Figure \ref{fig:mixtureProbs} are concentrated around the points 1, 14, 40 and 66, which match some of the regions highlighted on Figure \ref{fig:MDDMs} with the MDDM and the predictions on Figure \ref{fig:predHellingerDist}. Results when kriging is applied are again similar to those when it is not applied. Therefore, we can notice that the mixture method can also identify possible change points related to change times in the satellite images.

\section{Conclusion}

We propose nonparametric methods for analyzing satellite image time series through multi-date divergence matrices. The methods proposed take advantage of fast computations obtained with wavelet techniques and, using a functional time series approach, allow us to evaluate change predictions of the images. A kriging approach was also considered and provided an interesting way to pre-smooth the images, taking into account their spatial variation. The application of these methods to real data shows their feasibility even for large data, and the results are consistent with the experts analysis.

\newpage
\bibliographystyle{IEEEbib}
\bibliography{bibfile}

\end{document}